\documentclass[12pt]{article}

\usepackage{graphicx}

\newcommand{\beq}{\begin{equation}}
\newcommand{\eeq}{\end{equation}}
\newcommand{\beqa}{\begin{eqnarray}}
\newcommand{\eeqa}{\end{eqnarray}}
\newcommand{\beqar}{\begin{eqnarray*}}
\newcommand{\eeqar}{\end{eqnarray*}}

\begin{document}
\thispagestyle{empty}

\vspace{25pt}
\begin{center}
{\textbf{\Large A sterile neutrino at MiniBooNE and 
IceCube}}\footnote{Talk presented at 
{\it II Russian-Spanish Congress: Particle and
Nuclear Physics at all Scales}, Saint-Petersburg, October 1-4, 2013.}

\vspace{25pt}

Manuel Masip
\vspace{12pt}

\textit{CAFPE and Departamento de F{\'\i}sica Te\'orica y del Cosmos}\\ 
\textit{Universidad de Granada, 18071 Granada, Spain}\\
\vspace{15pt}
\texttt{masip@ugr.es}
\end{center}

\vspace{25pt}

\date{\today}

\begin{abstract}
We discuss the possibility that a sterile 
neutrino of mass around $50$ MeV slightly mixed with the muon
flavor may be the origin of the MiniBooNE 
anomaly. We show that its production in the atmosphere in a
fraction of kaon decays would imply an 
excess of contained showers at IceCube from down-going and
near-horizontal directions.
\end{abstract}

\newpage

\section{Introduction}

During the past 20 years a number of experiments with solar, 
atmospheric, reactor and baseline neutrinos have shown that 
neutrinos have masses and mixings \cite{Beringer:1900zz}. 
These experiments 
provide a framework with
\begin{equation}
\left\{ 
\begin{array}{l} 
\displaystyle 
\Delta m_{12}^2 \approx 7.6\times 10^{-5}\;{\rm eV^2}\\
\Delta m_{23}^2 \approx 2.4\times 10^{-3}\;{\rm eV^2}\\
\;\;\;\;\;\;\;\;\;\; \approx \Delta m_{13}^2 
\end{array} \right.
\;\;\;\;\;\;\;\;\;
\left\{ 
\begin{array}{l} 
\displaystyle 
\sin^2\theta_{12}\approx 0.30\\
\sin^2\theta_{23}\approx 0.50\\
\sin^2\theta_{13}\approx 0.025
\end{array} \right. 
\end{equation}
that fits remarkably well most of the data. 
From a model-building point of view 
the importance of this discovery cannot be overstated.
We should, however, keep in mind that
\begin{itemize}
\item
There are some very basic questions with no 
answer yet. We do not know, in particular, if these masses
are purely electroweak (EW) (just like the electron mass) or if
neutrinos are Majorana fermions and their mass is  
revealing a new scale in particle physics (like 
radiactivity reveals the EW scale):
\begin{equation}
 y_\nu \; H L \nu^c\;\;
{\rm or} \;\;
{1\over \Lambda_\nu}\;
H H L L\;?
\end{equation}
\item There has been a \emph{persistent} anomaly in 
several experiments with neutrino beams from particle
accelerators. Namely, LSND \cite{Athanassopoulos:1996jb} and 
MiniBooNE \cite{AguilarArevalo:2007it,AguilarArevalo:2009xn}
have observed an excess 
of $\approx 3$ events with an electron in the final state per each 
1000 $\nu_\mu$ charged-current (CC) interactions. The interpretation
of these events in terms of $\nu_\mu\to \nu_e$ oscillations would force 
the addition of two sterile neutrinos of mass around 
1 eV. Here we will 
go in a completely different direction: we will introduce 
a sterile mode, but much heavier and unstable.
Let us very briefly review these experimental anomalies and
the heavy neutrino proposal.
\end{itemize}

\section{Gninenko's 50 MeV neutrino at LSND}

In the mid 90's LSND \cite{Athanassopoulos:1996jb} observed an 
excess of electron-like 
events that were interpreted as
$\bar \nu_\mu\to \bar \nu_e$ oscillations followed by a CC interaction
giving an electron and a free neutron,
$\bar \nu_e p\to e^+ n$. In particular, they could see 
the Cherenkov light of the
electron plus a 2.2 MeV photon coming from the capture of the
neutron to form a deuteron. 
Given the LSND fluxes \cite{Athanassopoulos:1996jb}, 
\begin{figure}
\begin{center}
  \includegraphics[height=.55\linewidth]{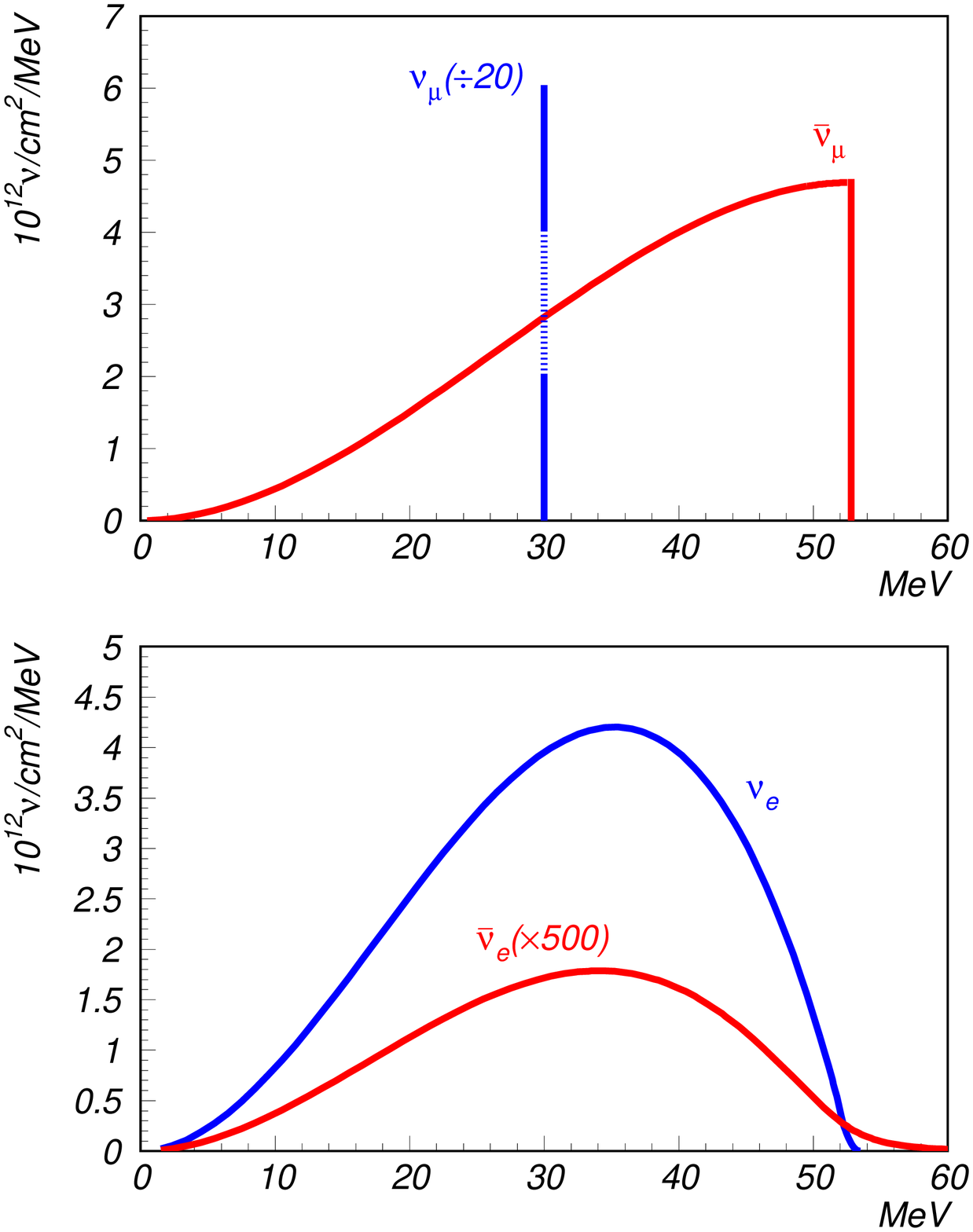}$\;\;\;$
  \includegraphics[height=.55\linewidth]{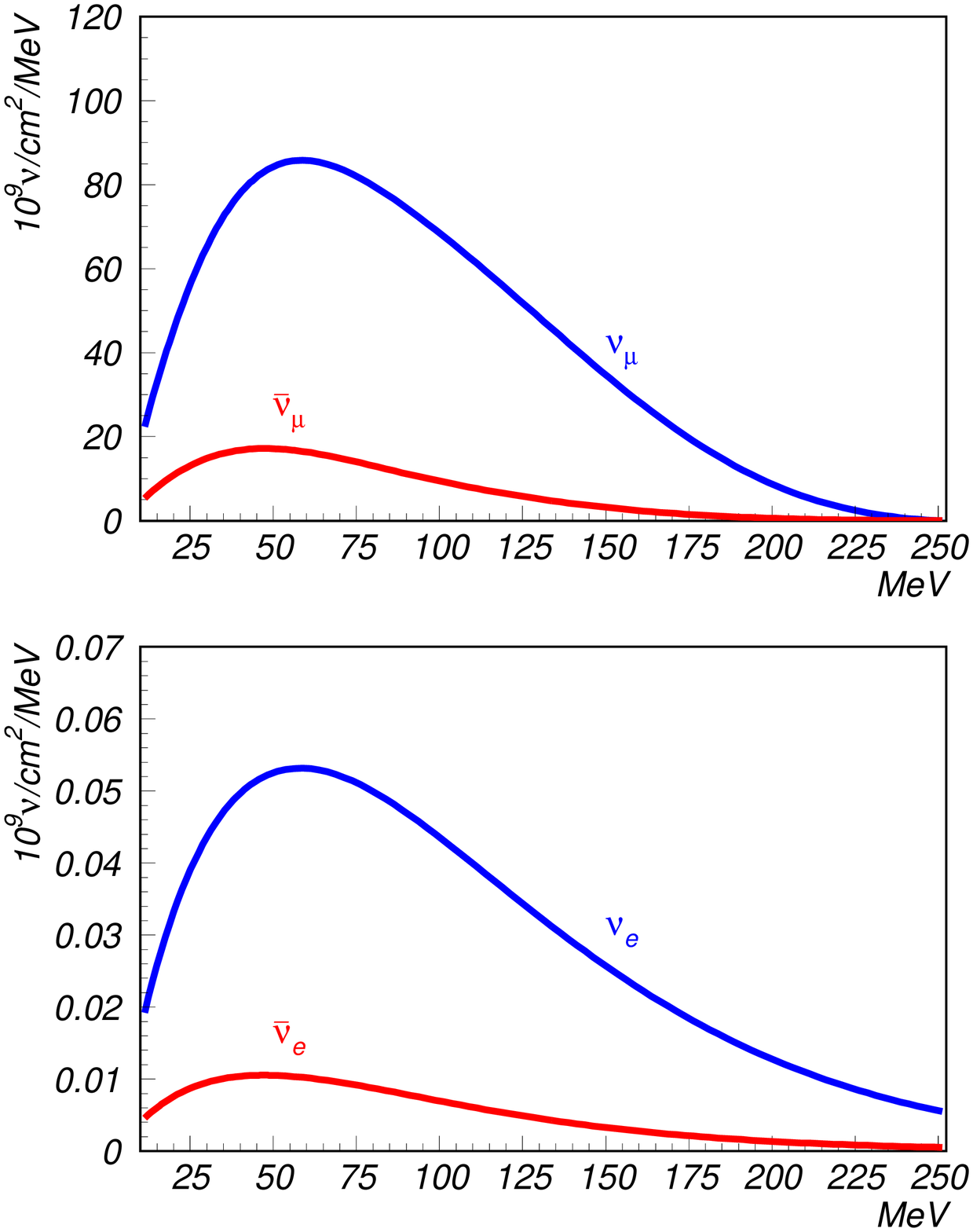}
\end{center}
  \caption{Neutrino fluxes at LSND \cite{Athanassopoulos:1996jb}.}
\end{figure}
the initial antineutrinos (in Fig.~1--left)
were coming from the decay-at-rest of positive muons, whereas
the appearence of oscillations would
require the existence of a $\approx 1$ eV sterile neutrino 
slightly mixed with the muon flavor. Instead,
Gninenko proposed three years ago a very different interpretation,
where the responsible for the signal would be the muon 
neutrinos from the decay-in-flight of pions (in Fig.~1--right), 
with energies of up to 200 MeV. 

Gninenko \cite{Gninenko:2010pr} showed that these events 
could be caused by an
interaction mediated by a Z boson (see Fig.~2)
that changes the initial $\nu_\mu$ 
into a sterile neutrino $\nu_h$ of mass around
50 MeV plus a free neutron. The neutrino should then decay
through an electromagnetic (EM) interaction
into a light neutrino plus a photon;
the photon would finally convert into an electron pair 
inside LSND giving the same Cherenkov ring as a 
single electron. In other words, the excess of electron-like 
events at LSND would actually be due to photon events.
For this to work the mixing of $\nu_h$ with the 
muon flavor must be relatively large, 
$|U_{\mu h}|^2\approx 10^{-3}$--$10^{-2}$,
and the EM dipole transition $\mu_{\rm tr}$
between $\nu_h$ and a light
neutrino must be such that 
the lifetime $\tau_h$ is shorter than $10^{-8}$ s.
\begin{figure}
\begin{center}
  \includegraphics[height=.15\linewidth]{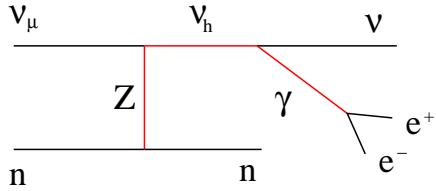}
\end{center}
  \caption{Basic process at LSND in Gninenko's scenario}
\end{figure}

The advantage of Gninenko's interpretation versus the oscillation
hypothesis is that it provides an explanation also for 
KARMEN \cite{Armbruster:2002mp}, that in 2002 tried to 
reproduce LSND using a similar 
technique and saw nothing. The initial
proton beam and the distances in both experiments were 
identical, but the larger angle of the neutrino beam 
selected at KARMEN ($90^o$, versus just $12^o$ at LSND)
implied a very different neutrino spectrum: 
KARMEN eliminated the neutrinos from 
in-flight decays, keeping only the less energetic antineutrinos
from $\mu^+$ decays at rest. 
Whereas oscillations should occur in both
experiments, the production of a 40--80 MeV neutrino
would be above threshold at KARMEN.

This mass and mixing 
of the the sterile neutrino may sound  \emph{unlikely};
is there really room for such a particle? Gninenko
\cite{Gninenko:2010pr} argues
that yes, that $\nu_h$ could appear instead of a regular 
$\nu_{\mu}$ in up to 1\% of all muon and kaon decays.
The bounds from current data depend on whether its lifetime is
longer or shorter than $10^{-9}$ s, 
which corresponds to a decay 
length of 30 cm. If the heavy neutrino is longer lived
then it tends to be invisible
({\it i.e.}, it does not decay inside the detector) and 
the kinematic changes that it
introduces in kaon and muon 
decays are not significant. 
If $\nu_h$ decays faster ({\it e.g.}, $\tau_h= 10^{-10}$ s) 
then the $K$ or $\mu$ decays will appear 
with an extra photon. It turns out, however, that 
this is also difficult to see due to background
processes with photons,
\begin{eqnarray}
&&{\rm BR}(\mu^- \to  e^-  \bar \nu_e \nu_\mu \gamma ) =
(1.4\pm 0.4)\% \nonumber \\
&&{\rm BR}( K^- \to  \mu^-  \bar \nu_\mu \gamma) =
(0.62\pm 0.08)\% \nonumber \\
&&{\rm BR}( K^- \to  \mu^- \bar \nu_\mu \pi^0) =
(3.35\pm 0.03)\%\,.
\end{eqnarray}
Although the limits on a 50 MeV neutrino $\nu_h$ with no decay modes
into charged particles are weaker than one may think, a recent 
analysis of the photon distribution in kaon decays 
by ISTRA+ \cite{Duk:2011yv}
disfavors lifetimes $\tau_h<10^{-9}$ s. These 
shorter lifetimes, however, are necessary in the original
Gninenko model in 
order to also explain the low-energy MiniBooNE 
anomaly (see below).
Moreover, it was noticed \cite{McKeen:2010rx}
that Gninenko's model is also in conflict
with observations of radiative muon capture at TRIUMF: the 
shorter-lived heavy neutrino is produced and decays 
too often there, implying three
times more events than observed. 

\section{A variation of the model at MiniBooNE}
We propose \cite{Masip:2012ke}
a variation of Gninenko's model with three
basic ingredients.
\begin{itemize}
\item
We consider a longer-lived $\nu_h$ ({\it e.g.}, 
$\tau_h=5\times 10^{-9}$ s). 
The target volume at TRIUMF 
is smaller than $c\tau_h$, and the 
number of events with $\nu_h$ decaying inside the detector
goes as $1/\tau_h$ when the lifetime grows. 
The bounds at ISTRA+ \cite{Duk:2011yv}
also weaken with an increased $\tau_h$.
\item
We include the EM production of $\nu_h$ (see Fig.~3).
\begin{figure}
\begin{center}
  \includegraphics[height=.18\linewidth]{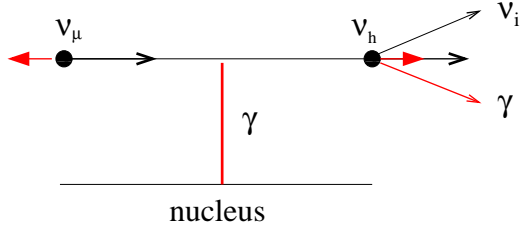}
\end{center}
  \caption{Photon-mediated process at MiniBooNE \cite{Masip:2012ke}.}
\end{figure}
The same EM operator that lets $\nu_h$ decay into $\gamma \nu_i$
implies,
when the light neutrino $\nu_i$ is a $\nu_\mu$, a production 
channel at MiniBooNE that was overlooked by Gninenko. The
extra $\nu_h$ produced through photon exchange will dominate
and compensate
the reduction due to the increased lifetime at MiniBooNE,
as some $\nu_h$ will decay outside the detector. 
If we write
\begin{equation}
L_{eff}\supset {1\over 2} \, \mu_{\rm tr}^{ih} \,
\Big( \overline \nu_h\, \sigma _{\mu \nu}
 \left(1-\gamma_5\right) \nu_i + \overline \nu_i\, \sigma _{\mu \nu}  
\left(1+\gamma_5\right) \nu_h \Big) \, \partial^\mu A^\nu\,,
\end{equation}
a lifetime $\tau_h=5\times 10^{-9}$ s implies 
$(\sum_i(\mu_{\rm tr}^{ih})^2)^{1/2}=7\times 10^{-6}\;{\rm GeV}^{-1}=
2\times 10^{-8}\mu_B$, whereas the anomaly at 
MiniBooNE  will require 
$\mu_{\rm tr}^{\mu h}=2\times 10^{-9}\mu_B$. This means that only 1\%
of $\nu_h$ decays go into a muon neutrino.
\item
Finally, we 
impose \cite{Masip:2012ke} that the $\nu_h$ is a Dirac particle:
neutrino plus antineutrino of both chiralities. This is important
because it will reduce the energy of the events at MiniBooNE.
Basically, the initial $\nu_\mu$ 
of negative helicity becomes, after the EM transition,
a heavy neutrino with the spin pointing 
($99.9\%$ of the times) forward. It turns out that the 
decay is then not isotropic and the photon is preferably emited 
backwards, with less energy than if it were emited forward or 
isotropically. The Dirac nature and the increased 
lifetime that we have
assumed are essential in order to generate an excess only at
low energies (higher-energy neutrinos tend to decay outside the
detector). 
\end{itemize}

Just like LSND, MiniBooNE is unable to distinguish an
electron from a photon converted into an $e^+ e^-$ pair.
Our estimate \cite{Masip:2012ke} of the number of photon events from $\nu_h$
decays is given in Fig.~4 for $m_h=50$ MeV, 
$|U_{\mu h}|^2= 0.003$, $\tau_h=5\times 10^{-9}$ s and
$\mu_{\rm tr}^{\mu h}=2\times 10^{-9}\mu_B$. 
On the left, we plot 
\begin{figure}
\begin{center}
  \includegraphics[height=.38\linewidth]{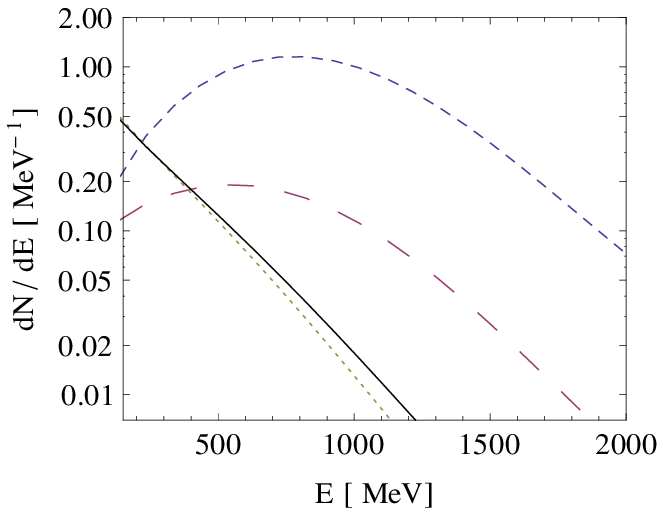}$\;\;\;$
  \includegraphics[height=.367\linewidth]{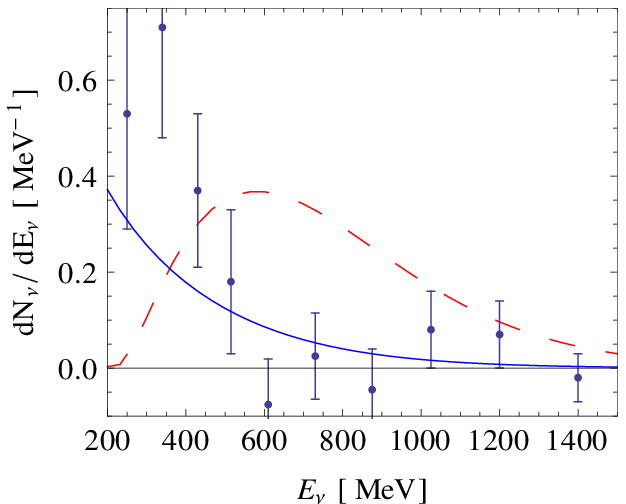}
\end{center}
  \caption{Heavy neutrino events at MiniBooNE in the neutrino mode
($5.58\times 10^{20}$ POT) \cite{Masip:2012ke}.}
\end{figure}
with dashes the distribution of 
heavy neutrinos produced inside
the detector and with long dashes the distribution of 
neutrinos that decay inside the 
detector. Notice that these two lines are closer 
together at low energies and separate as the
energy grows. The dotted line is the
energy distribution of the photons (emitted preferably backwards)
from the decaying $\nu_h$,
and the final solid line is the
energy assigned to an initial $\nu_e$ 
that after a quasielastic collision becomes an electron 
with the same energy as the photon 
({\it i.e.}, the reconstruction of the photon events as 
$\nu_e$ events).

On the right, we plot the excess observed by MiniBooNE 
 with error
bars (zero would be consistency with the background) and in
dashes what could be expected from $\nu_\mu\to \nu_e$ oscillations
due to the 1 eV sterile 
neutrino that explains LSND. The initial publications by 
MiniBooNE emphasized that their results disprove the LSND 
oscillation hypothesis, although in the later ones they 
fail to explain the low-energy excess observed in the
data \cite{AguilarArevalo:2007it,AguilarArevalo:2009xn}.
Our heavy neutrino hypothesis gives a 
remarkable\footnote{We do not consider significant the different
angular distribution of these photon-mediated events  
discussed in \cite{Radionov:2013mca}}
fit to the
data, with similar results for the MiniBooNE excess in the
antineutrino mode in Fig.~5.
\begin{figure}
 \begin{center}
 \includegraphics[height=.39\linewidth]{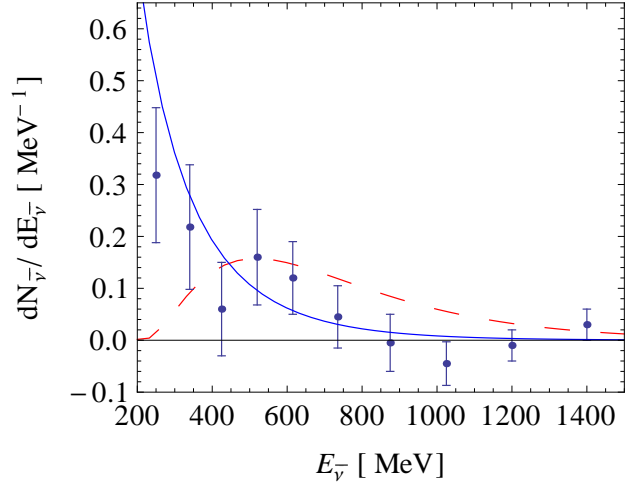}
\end{center}
  \caption{Heavy neutrino events at MiniBooNE in the 
antineutrino mode ($11.27\times 10^{20}$ POT) \cite{Masip:2012ke}.}
\end{figure}

Three coments are here in order.
\begin{itemize}
\item
The magnetic dipole transitions that we assume can be 
generated \cite{Bueno:2013mfa}
in left-right symmetric completions of the model 
at the TeV scale. In particular, we just need that both the
left and the righ-handed components of $\nu_h$ define 
$SU(2)_R$ doublets together with a charged lepton, and that
the breaking of this gauge symmetry gives large mass to 
this charged lepton while keeping $\nu_h$ light.
\item
An analysis of T2K (where most of the initial
muon neutrinos have oscillated into the
tau flavor at Super-Kamiokande) data 
\cite{Abe:2011sj}
shows that the decay $\nu_h\to \gamma \nu_{\tau}$ 
has a $< 1\%$ branching 
ratio \cite{Masip:2012ke}. Therefore,
there must be a second sterile neutrino $\nu_{h'}$ lighter,
longer lived and less mixed with the standard flavors 
than $\nu_h$
that accounts for $99\%$ of the decays: $\nu_h\to
\gamma \nu_{h'}$.
 \item
MicroBooNE \cite{Jones:2011ci}
will investigate whether the low-energy
excess at MiniBooNE is caused by electron or by 
photon events. It has been argued \cite{Hill:2010zy}
that the standard
background producing a photon (in Fig.~6) may 
be underestimated in current simulations.
Hopefully, there are observables that may distinguish this 
background from the $\nu_h\to \gamma \nu_{h'}$ hypothesis
(in particular, the 
longitudinal event distribution inside the detector is 
flat for the background 
but proportional to $(1-e^{-z/\lambda_d})\approx {z\over \lambda_d}$ 
in $\nu_h$ events), so the MiniBooNE puzzle should 
be settled during the next year.
\end{itemize}
\begin{figure}
\begin{center}
  \includegraphics[height=.2\linewidth]{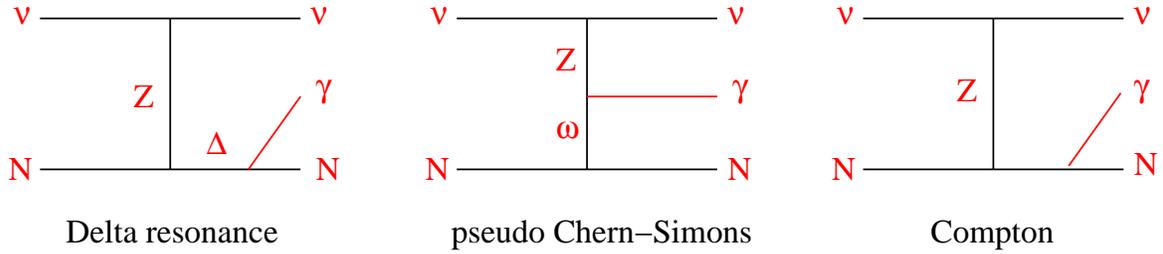}
\end{center}
  \caption{Some background processes producing a photon at MiniBooNE.}
\end{figure}

\section{Implications at IceCube}
The LSND and the MiniBooNE anomalies consist of
an excess of events with an electron in the final state. The 
question now would be, could one expect (at larger energies and 
distances) a similar 
excess in neutrino telescopes? 
Would it be detectable?

A telescope like IceCube observes two types of events:
muon tracks and point-like energy depositions (hadronic 
and EM showers 
develop in $\approx 10$ m of ice) \cite{Halzen:2010yj}. 
The direction
of the first ones can be determined very precisely, 
while in cascade events the error is around $\pm 15^o$.
Upgoing muons crossing the telescope come necesarily 
from $\nu_\mu$ interactions (most of these tracks do not 
start inside the detector), whereas contained cascades from any direction
correspond to $\nu_{e,\tau}$ CC interactions
or $\nu_{e,\mu,\tau}$ neutral current (NC) interactions.

At energies above 100 GeV the flux of atmospheric 
muons and neutrinos is dominated 
by kaon decays \cite{Lipari:1993hd}. 
Since our
$\nu_h$ appears in up to $\approx 1\%$ of these decays instead
of a muon neutrino, it will be abundant in the atmosphere.
Its decay length ($\gamma c \tau_h$) at TeV energies 
becomes larger than 10 km, so 
an atmospheric $\nu_h$ 
could reach the center of IceCube 
and decay there. The resulting photon would produce a 
shower-like event similar to a CC $\nu_e$ event or to 
an inelastic NC collision.

\begin{figure}
\begin{center}
  \includegraphics[height=.38\linewidth]{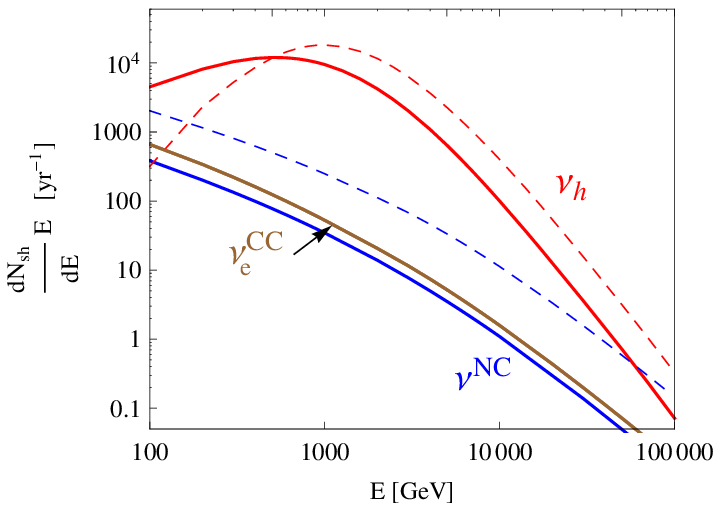}
$\;\;\;\;\;\;\;\;\;\;\;\;\;\;\;$
  \includegraphics[height=.22\linewidth]{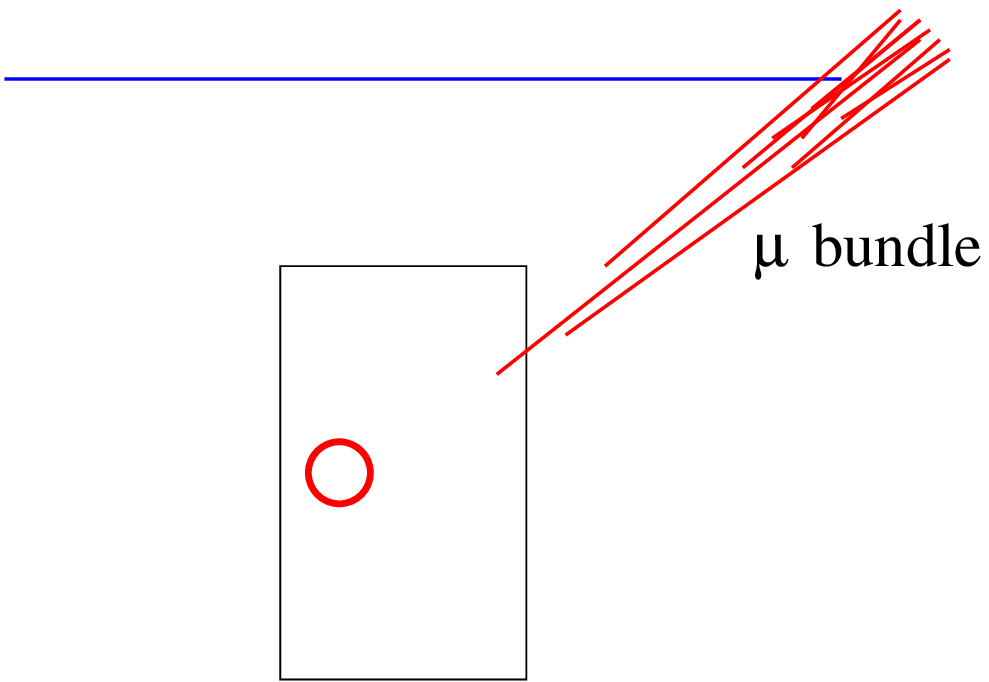}
\end{center}
  \caption{Cascade events at a neutrino telescope from
$\nu_e$ CC interactions, $\nu_{e,\mu}$ NC interactions, and 
$\nu_h\to \gamma \nu_i$ decays \cite{Masip:2011qb}. Neutrinos
from charmed-hadron decays, dominant at $E\ge 10^5$ GeV, have
not been included. In dashes, the energy of the parent neutrino (left).
Typical $\nu_h$ event at IceCube (right).}
\end{figure}
We have investigated these processes \cite{Masip:2011qb}
and have obtained that $\nu_h$ decays 
may change substantially the number of contained 
cascade-like events between 100 GeV (at lower
energies the heavy neutrino can not reach the telescope)
and 1 PeV (at very high energies the decay length grows 
and the probability to decay inside the telescope becomes
negligible). In Fig.~7
we have used the Z-moment method to estimate their
frequency. 
Generically, the excess of TeV contained 
showers introduced by the heavy neutrino has 
two basic features:
\begin{itemize}
\item
The excess would only appear in down-going or near-horizontal 
events (there are no $\nu_h$ upgoing events)
\item
Some of these events (specially the ones from small 
zenith angles) would be contaminated with muons. Since
the heavy neutrino is produced in the atmosphere together
with a very energetic muon, there will be an 
excess of \emph{muon plus contained cascade} events.
\end{itemize}

We would like to conclude by noticing two recent
observations published by IceCube. The first one is the 
sample of 
28 events above 30 TeV described in \cite{Aartsen:2013jdh}, which are
indeed very interesting data.
If we discount 4 muon-like tracks that are consistent
with atmospheric muons entering the telescope, we are
left with 24 events that should correspond to neutrino 
interactions. Three of these events are muons and 21
are cascades. The 3/21 ratio suggests that these events
are not atmospheric in origin: below 100 TeV the lepton fluxes
are still dominated by kaon and pion decays, which produce
more muons than electrons (atmospheric taus are 
irrelevant). 
There are just too many cascades relative to muons 
to be consistent with standard
atmospheric-neutrino interactions.
The energy distribution of 
these events is also much flatter ($\approx E^{-2}$)
than the one from atmospheric neutrinos ($\approx E^{-3.7}$). 
Therefore, it is very likely that
this is the first observation of cosmic neutrinos. On the
other hand, it is \emph{intriguing} that all the
cascade events of energy below 100 TeV 
are down-going or near-horizontal, since at these energies
the Earth is not fully opaque (from zenith angles between 
$90^o$ and $150^o$) to neutrinos, and the events
observed do not show preference for the galactic disk neither.
In our opinion, the data implies an excess of down-going cascade
events relative to up-going cascades if one 
assumes a cosmic origin or relative to muon events if one assumes
a standard atmospheric origin, 
and it could be interesting to
run a simulation of the atmospheric $\nu_h$ hypothesis.

\begin{figure}
\begin{center}
  \includegraphics[height=.3\linewidth]{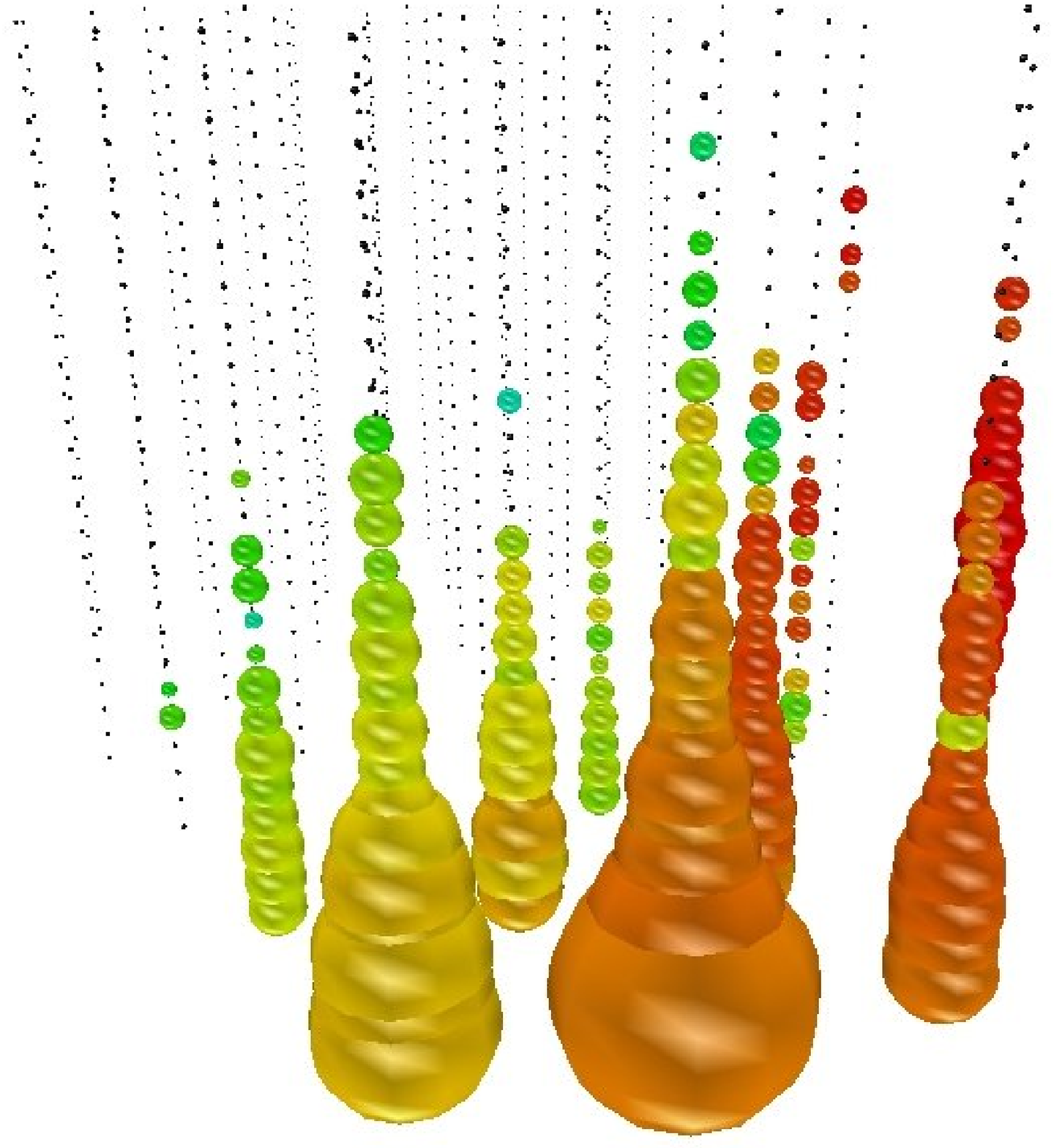}
$\;\;\;\;\;\;\;\;$
  \includegraphics[height=.3\linewidth]{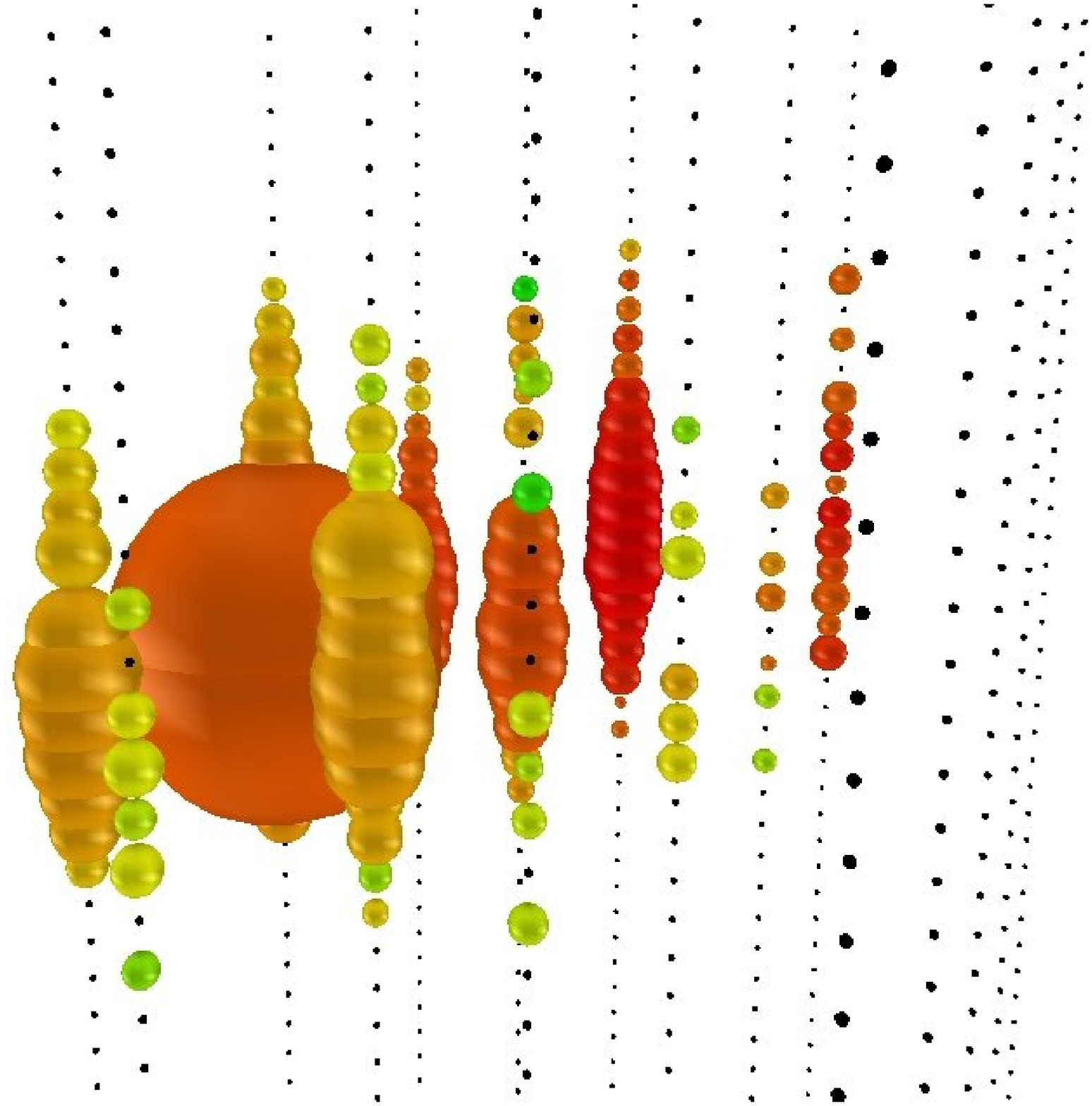}
\end{center}
  \caption{High Light Density events selected in \cite{posselt:2013}.}
\end{figure}
A second interesting observation by IceCube 
\cite{posselt:2013} has been 
obtained in the search for High Light Density events 
related to magnetic monopoles and other exotics.
Although they did not find any monopole candidate, 
but their cuts selected two events (in Fig.~8) for an estimated 
background of 0.14 events. It seems to us that these 
two events are not regular contained showers
nor muon bundles, but that they could be interpreted 
as a cascade event \emph{inside} a muon bundle. Such
topology would be difficult to explain with standard
model particles, but it is exactly what the heavy neutrino 
hypothesis would suggest.

\section{Summary and discussion}

Neutrino physics has progressed a great deal 
during the past 20 years, but a few
basic questions and some persistent anomalies
should still be clarified. We have focused on MiniBooNE and
IceCube, that have provided very interesting although 
still incomplete results. Hopefully, in the near
future MicroBooNE
and an increased statistics, respectively, will 
establish the reach of their current results.

We have argued that there is room in neutrino 
physics for \emph{serious} departures
from the three-flavor picture, 
in particular, we have discussed a 50 MeV 
sterile mode that could be a solution to 
the baseline anomalies.

We find the possible incidence of such neutrino in
completely different experiments, like neutrino telescopes,
very interesting.
The heavy neutrino could be produced in the atmosphere
and decay inside the telescope, introducing an excess of contained
showers from downgoing and near horizontal directions,
some of them contaminated by atmospheric muons. There 
seems to be room in current IceCube 
data for this type of events. Telescopes have the 
advantage that they are basically background free experients: 
each individual event must be explained in terms of neutrino
interactions or atmospheric muons. In this sense, 
the 50 MeV neutrino 
is a possibility that that may be worth exploring.

\section*{Acknowledgments}
This work has been partially supported by
MICINN of Spain (FPA2010-16802 and 
Consolider-Ingenio 
{\bf Multidark} CSD2009-00064), and   
by Junta de Andaluc\'{\i}a
(FQM 101 and FQM 3048).


\begin{thebibliography}{99}


\bibitem{Beringer:1900zz}
  J.~Beringer {\it et al.}  [Particle Data Group Collaboration],
  Phys.\ Rev.\ D {\bf 86} (2012) 010001.

\bibitem{Athanassopoulos:1996jb}
  C.~Athanassopoulos {\it et al.}  [LSND Collaboration],
  Phys.\ Rev.\ Lett.\  {\bf 77} (1996) 3082;
  [arXiv:nucl-ex/9605003];
  C.~Athanassopoulos {\it et al.}  [LSND Collaboration],
  Phys.\ Rev.\  C {\bf 54} (1996) 2685;
  [arXiv:nucl-ex/9605001];
  A.~Aguilar {\it et al.}  [LSND Collaboration],
  Phys.\ Rev.\  D {\bf 64} (2001) 112007.
  [arXiv:hep-ex/0104049].

\bibitem{AguilarArevalo:2007it}
  A.~A.~Aguilar-Arevalo {\it et al.}  [The MiniBooNE Collaboration],
  Phys.\ Rev.\ Lett.\  {\bf 98} (2007) 231801;
  [arXiv:0704.1500 [hep-ex]].
  A.~A.~Aguilar-Arevalo {\it et al.}  [MiniBooNE Collaboration],
  Phys.\ Rev.\ Lett.\  {\bf 102} (2009) 101802.
  [arXiv:0812.2243 [hep-ex]].

\bibitem{AguilarArevalo:2009xn}
  A.~A.~Aguilar-Arevalo {\it et al.}  [MiniBooNE Collaboration],
  Phys.\ Rev.\ Lett.\  {\bf 103} (2009) 111801
  [arXiv:0904.1958 [hep-ex]];
  A.~A.~Aguilar-Arevalo {\it et al.}  [MiniBooNE Collaboration],
  Phys.\ Rev.\ Lett.\  {\bf 105} (2010) 181801
  [arXiv:1007.1150 [hep-ex]].

\bibitem{Gninenko:2010pr}
  S.~N.~Gninenko,
  Phys.\ Rev.\  D {\bf 83} (2011) 015015.
  [arXiv:1009.5536 [hep-ph]].

\bibitem{Armbruster:2002mp}
  B.~Armbruster {\it et al.}  [KARMEN Collaboration],
  Phys.\ Rev.\  D {\bf 65} (2002) 112001.
  [arXiv:hep-ex/0203021].

\bibitem{Duk:2011yv}
  V.~A.~Duk {\it et al.}  [ISTRA+ Collaboration],
  Phys.\ Lett.\ B {\bf 710} (2012) 307
  [arXiv:1110.1610 [hep-ex]].

\bibitem{McKeen:2010rx}
  D.~McKeen and M.~Pospelov,
  Phys.\ Rev.\  D {\bf 82} (2010) 113018.
  [arXiv:1011.3046 [hep-ph]].

\bibitem{Masip:2012ke}
  M.~Masip, P.~Masjuan and D.~Meloni,
  JHEP {\bf 1301} (2013) 106
  [arXiv:1210.1519 [hep-ph]].

\bibitem{Radionov:2013mca}
  A.~Radionov,
  Phys.\ Rev.\ D {\bf 88} (2013) 015016
  [arXiv:1303.4587 [hep-ph]].

\bibitem{Bueno:2013mfa}
  A.~Bueno, M.~Masip, P.~Sánchez-Lucas and N.~Setzer,
  Phys.\ Rev.\ D {\bf 88} (2013) 073010
  [arXiv:1308.0011 [hep-ph]].

\bibitem{Abe:2011sj}
  K.~Abe {\it et al.}  [T2K Collaboration],
  Phys.\ Rev.\ Lett.\  {\bf 107} (2011) 041801
  [arXiv:1106.2822 [hep-ex]].

\bibitem{Jones:2011ci}
  B.~J PJones,
  PoS EPS {\bf -HEP2011} (2011) 436
   [J.\ Phys.\ Conf.\ Ser.\  {\bf 408} (2013) 012028]
  [arXiv:1110.1678 [physics.ins-det]].

\bibitem{Hill:2010zy}
  R.~J.~Hill,
  Phys.\ Rev.\ D {\bf 84} (2011) 017501
  [arXiv:1002.4215 [hep-ph]].

\bibitem{Halzen:2010yj}
  F.~Halzen and S.~R.~Klein,
  Rev.\ Sci.\ Instrum.\  {\bf 81} (2010) 081101.


\bibitem{Lipari:1993hd}
  P.~Lipari,
  Astropart.\ Phys.\  {\bf 1} (1993) 195;
  J.~I.~Illana, P.~Lipari, M.~Masip and D.~Meloni,
  Astropart.\ Phys.\  {\bf 34} (2011) 663.

\bibitem{Masip:2011qb}
  M.~Masip and P.~Masjuan,
  Phys.\ Rev.\ D {\bf 83} (2011) 091301
  [arXiv:1103.0689 [hep-ph]].

\bibitem{Aartsen:2013jdh}
  M.~G.~Aartsen {\it et al.}  [IceCube Collaboration],
  Science {\bf 342} (2013) 6161,  947
  [arXiv:1311.5238 [astro-ph.HE]].

\bibitem{posselt:2013}
J.~Posselt, {\it Search for Relativistic Magnetic Monopoles
with the IceCube 40-String Detector}, Ph.D. Dissertation, Wuppertal
Univ., October 2013 
(http://elpub.bib.uni-wuppertal.de/servlets/DocumentServlet?id=3814); 
  R.~Abbasi {\it et al.}  [IceCube Collaboration],
  Phys.\ Rev.\ D {\bf 87} (2013) 022001
  [arXiv:1208.4861 [astro-ph.HE]].


\end{thebibliography}
\end{document}